\documentclass[aps,amsmath,amssymb,superscriptaddress,twocolumn,showkeys,showpacs]{revtex4}

\usepackage{graphicx}

\begin{document}

\title{A fretting crack initiation prediction taking into account the surface roughness and the crack nucleation process volume}

\author{H. Proudhon}
\affiliation{Laboratoire de Tribologie et de dynamique des syst\`{e}mes, UMR 5513, Ecole Centrale de Lyon, 36~Avenue Guy de Collongue, 69134 Ecully Cedex, France}
\affiliation{Groupe d'Etude de M\'{e}tallurgie et de Physique des Mat\'{e}riaux, UMR 5510, Insa de Lyon, 20~avenue Albert~Einstein, 69621 Villeurbanne Cedex, France}

\author{S. Fouvry}
\thanks{Corresponding author. Tel.: +33-472-186-562; fax: +33-472-433-383.}
\email{siegfried.fouvry@ec-lyon.fr}
\affiliation{Laboratoire de Tribologie et de dynamique des syst\`{e}mes, UMR 5513, Ecole Centrale de Lyon, 36~Avenue Guy de Collongue, 69134 Ecully Cedex, France}

\author{J-Y. Buffi\`{e}re}
\affiliation{Groupe d'Etude de M\'{e}tallurgie et de Physique des Mat\'{e}riaux, UMR 5510, Insa de Lyon, 20~avenue Albert~Einstein, 69621 Villeurbanne Cedex, France}
\begin{abstract}
This paper presents an experimental study of the fretting crack nucleation threshold, expressed in terms of loading conditions, with a cylinder/plane contact. The studied material is a damage tolerant aluminium alloy widely used in the aerospace application. Since in industrial problems, the surface quality is often variable, the impact of a unidirectional roughness is investigated via varying the roughness of the counter body in the fretting experiments. As expected, experimental results show a large effect of the contact roughness on the crack nucleation conditions. Rationalisation of the crack nucleation boundary independently of the studied roughnesses was successfully obtained by introducing the concept of effective contact area. This does show that the fretting crack nucleation of the studied material can be efficiently described by the local effective loadings inside the contact. Analytical prediction of the crack nucleation is presented with the Smith-Watson-Topper (SWT) parameter and size effect is also studied and discussed.
\end{abstract}


\maketitle
\section{Introduction}\label{sec:intro}
Weight saving is still a topical issue in aerospace applications. Because of their low densities, aluminium alloys are widely used in air plane structures, in spite of quite bad fatigue and tribological properties. As noted by many authors \cite{Szolwinski1996,Harish1997}, a very important example of application is riveted lap joints structures, where contact fretting stress, due to vibrations during the flight, can initiate cracks at the hole edge which may propagate due to the overall structure fatigue stress. In the worst case, propagation will lead to failure. Authors \cite{Harish1997,Kapsa2003} also report partial slip in rivet structures. This shows the importance of investigating fretting crack initiation boundary in partial slip conditions of concerned material. In addition, in industrial application, where variable surface qualities can be found, one major issue is the impact of surface roughness on the crack nucleation process. Coupling experiments and modelling this paper focuses on this topic in order to predict the cracking risk on a 2024T351 aluminium alloy under fretting loading and taking into account the roughness impact. To evaluate the roughness effect on a simple fretting test, a cylinder/plane configuration with three different fretting pads in terms of roughness, was used. After a brief description of fretting test, material studied and surface roughness morphologies, the crack nucleation boundary at $5.10^4$ fretting cycles under partial slip condition is defined for the smooth contact case. Two unidirectional surface roughnesses are successively analysed and the corresponding crack nucleation boundary is determined. Introducing the concept of effective contact area, the different experimental results are rationalised through the introduction of a crack nucleation master curve. Finally, applying the SWT multiaxial fatigue criterion, prediction of the crack nucleation is intended taking into account the size effect.
\section{Experimental work}\label{expe.sec}
Fretting is defined as a small oscillatory movement between two bodies in contact which induce a relative displacement between 
the two surfaces. Considered as a plague for modern industries, it is a very complex problem involving many aspects as tribology, contact fatigue mechanics and material science. To reproduce such phenomena, different fretting test devices have been developed \cite{Nowell1990,Blanchard1991}.
\subsection{Material and specimens description}
\begin{figure}[!hbt]
\begin{center}
\includegraphics[width=75mm]{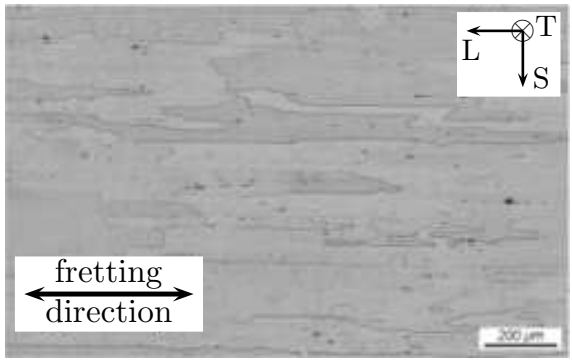}
\end{center}
\caption{Al 2024T351 microstructure of the sample surface. Optical micrograph, after a Keller reactive attack, showing elongated grain structure in the rolling direction. Intermetallics particles appear in dark.}
\label{microstructure.fig}
\end{figure}

2024T351 is an aluminium copper magnesium alloy which has been solution heat treated, control stretched and naturally aged. 
Chemical composition and mechanical properties are listed in tables~\ref{compo.tab} and~\ref{mecha.tab}, respectively. Rp$_{0.2\%}$ denotes the classical yield stress defined at 0.2\% of strain ; E the young modulus, $\nu$ denotes the poisson's ration and $\sigma_d$ the fatigue limit.
\begin{table}[!hbt]
\caption{Chemical composition of Al 2024T351 (weight \%)}
\begin{tabular}{ccccccc}
	\%Cu & \%Mg & \%Mn & \%Fe & \%Si & \%Cr & \%Ti \\
	\hline
	4.4 & 1.45 & 0.62 & 0.20 & 0.15 & 0.01 & 0.03 \\
\end{tabular} 
\label{compo.tab}
\end{table}
\vspace{0.5cm}
\begin{table}[!hbt]
\caption{A summary of the mechanical properties of the 2024T351 alloy used for fretting tests.}
\label{mecha.tab}
\begin{tabular}{cccc}
	Rp$_{0.2\%}$(MPa) & E(GPa) & $\nu$ &  $\sigma_d$(MPa) \\
	\hline
	325 & 72.4 & 0.33 & 140 \\
\end{tabular} \\
\end{table}
Specimens were extracted from a laminar plate of 25 mm thick supplied by Pechiney Industries. After machining, the surface in contact were carefully polished until mirror state ($R_a\mbox{ around }0.05 \mu m$).Displacement direction of the fretting tests is parallel to the L direction of the material along which grains are elongated(see fig.~\ref{microstructure.fig}). Grain size in the T direction was determined to be $150 \mu m$ by Electron Back Scattered Diffraction (EBSD) mapping, using a line intercept measurement method.
\subsection{Fretting experiment}\label{fretting_exp.ssc}
\begin{figure}[!hbt]
\begin{center}
\includegraphics[width=75mm]{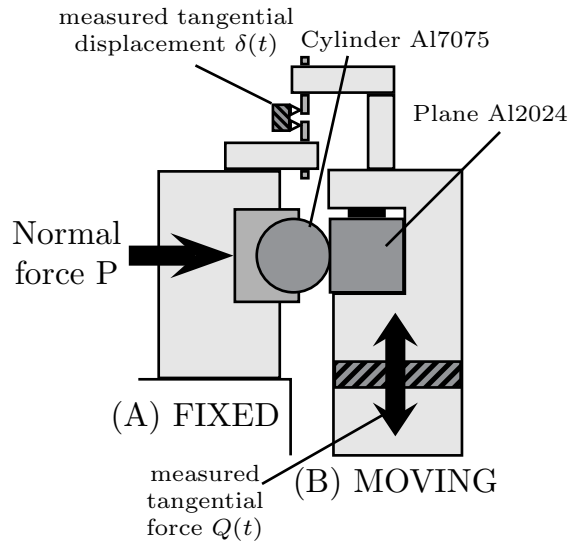}
\end{center}
\caption{Detail of fretting test device.}
\label{montage.fig}
\end{figure}

\begin{figure}[!hbt]
\begin{center}
\includegraphics[width=75mm]{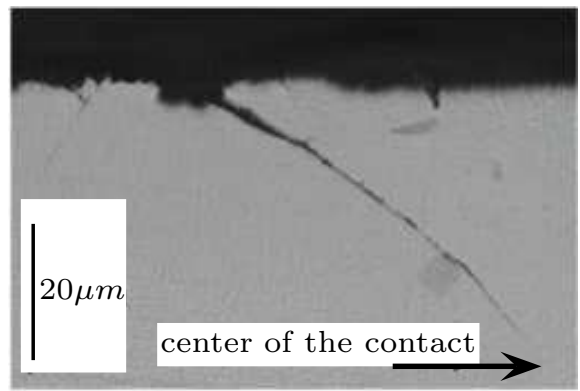}
\caption{Optical micrograph of the cross section below the surface.}
\label{crack.fig}
\end{center}
\end{figure}

The experimental setup used in this study is based on a fretting device rigidly mounted to a servo hydraulic test machine (see figure~\ref{montage.fig}). The counter body is made of 7075T6 aluminium alloy ; the cylinder radius is 49 mm and the cylinder length $L$ is 4.4 mm. In cylinder plane configuration, it is helpful to define normalised loads $P$ and $Q$ with respect to the contact length (eq.~\ref{PQ.eqn}), which is perpendicular to the sliding direction. The linear normal load range is 100 to 1000 N/mm, equivalent to a nominal maximum hertzian pressure from 220 MPa to 700 MPa.
\begin{equation}
P=\frac{F_N}{L}\quad{}\mbox{and}\quad{}Q=\frac{F_T}{L}
\label{PQ.eqn}
\end{equation}
Once specimen is fixed and carefully aligned with the counter body, fretting solicitation is applied during $5.10^4$ cycles by monitoring the normal load imposed to the surfaces ($P$) and the displacement amplitude ($\delta$) or the tangential load amplitude induced by the contact ($Q$). For the oscillatory parameters the stared versions ($\delta^*,Q^*$) denote their amplitude. After the test, an optical micrograph of the fretting scar is recorded and a cross section is made by cutting the specimen in the middle of the contact zone and polishing this surface. Crack initiation is investigated through optical micrography and images are recorded if any damage is visible. The crack angle with respect to the surface, and then crack depth below the surface are measured. Such a cross section of a fretting crack observed by optical micrography is presented in figure~\ref{crack.fig}.
\subsection{Surface quality}
\begin{figure}[!hbt]
\begin{center}
\includegraphics[width=75mm]{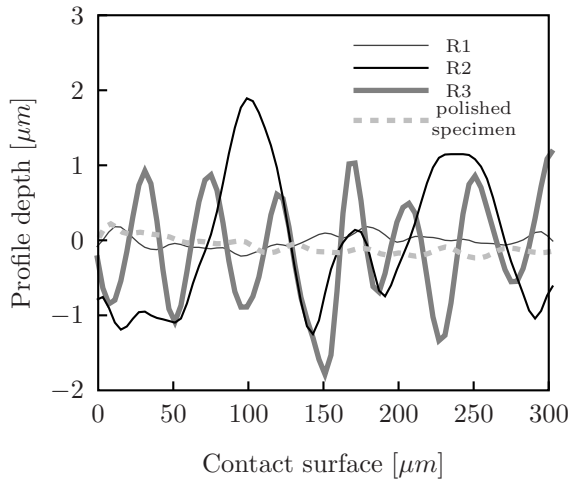}
\caption{Comparison of the 3 fretting pads roughness profiles.}
\label{profiles.fig}
\end{center}
\end{figure}

Relatively few literature can be found on the effect of roughness in fretting damage. Although a detrimental effect of the roughness is expected, some authors have reported a great benefit in term of lifetime can be obtain by a special laser surface texturing \cite{volchok2002}. In order to study the roughness effect on the fretting crack nucleation conditions, three different cylinders named R1, R2, R3 with increasing roughness, are used. Each counter body is obtained by lathe machining so that roughness is assumed unidimensional. R3 is machined with relatively high speed advance of the tool, R2 is obtained with a lower one and R1 is polished on the machine after as fine as possible machining. Figure~\ref{profiles.fig} presents the three profiles of the fretting pads and one of a specimen for comparison.
Respective roughnesses are measured on a tactile profilometer and the corresponding 
results are gathered in table~\ref{roughness.tab}.
\begin{table}[!hbt]
\caption{Mean and maximal roughness of cylinders and specimens}
\begin{tabular}{ccccc}
	& R1 & R2 & R3 & Samples \\
	\hline
	Ra($\mu $m) & 0.11 & 0.60 & 0.75 & 0.04-0.08 \\
	Rt($\mu $m) & 1.05 & 3.10 & 3.15 & 0.082 \\
\end{tabular}
\label{roughness.tab}
\end{table}
\section{Tribology analysis}\label{tribo.sec}
\subsection{Fretting regime analysis and material response maps}
\begin{figure}[!hbt]
\begin{center}
\includegraphics[width=75mm]{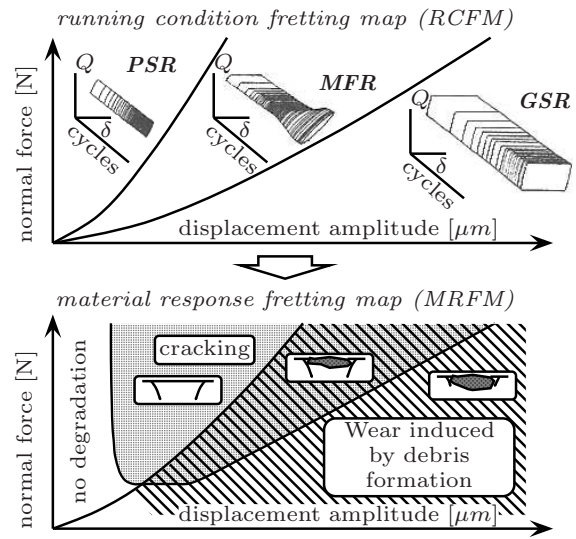}
\caption{Running condition fretting map which combines the fretting regime analysis (RCFM) with the material response (MRFM).}
\label{RCFM.fig}
\end{center}
\end{figure}

Two essential sliding conditions can be defined: Partial slip regime (PSR) characterised by a closed elliptical fretting loop, with a composite contact of a sliding and a sticking zone ; and Gross slip regime (GSR) identified by a quadratic dissipative fretting loop, with full sliding occurring over the entire contact zone \cite{Fouvry2003}. The transition between one regime to another defines the Mixed fretting regime (MFR). Identification of this regimes in carried out by plotting a Running Condition Fretting Map (RCFM). Waterhouse \cite{Waterhouse1981} first highlighted a correlation between the sliding regime and fretting damage. The main fretting damages can be rationalised in a Material Response Fretting Map (MRFM) involving non-degradation, cracking, particle detachment in a map with normal load versus relative displacement. The fretting description proposed by Vincent et al. \cite{Vincent1992} is based on these two sets of fretting maps (see figure~\ref{RCFM.fig}). When the relative displacement of the surface is small, partial slip prevail and induce crack initiation, although at higher values of the displacement, gross slip settles and impose wear of the surface. Since in our case, partial slip is assumed, very small displacement amplitudes have been maintained to study the crack initiation.
\subsection{Experimental fretting regimes determination}\label{exp_slid.sss}
\begin{figure}[!hbt]
\begin{center}
\includegraphics[width=75mm]{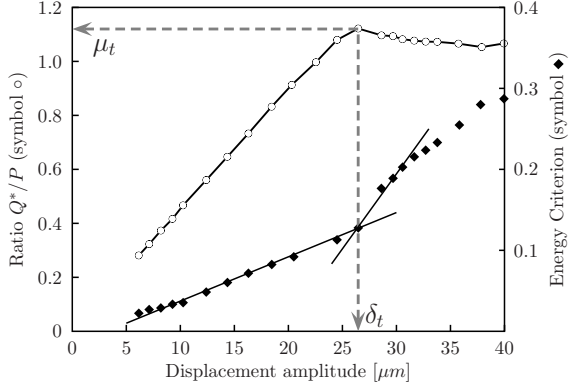}
\caption{Illustration of variable displacement method (see the text for details), Fn=750N. The ratio $Q^*/P$ (left scale) increases until stabilised gross slip condition is reach. The energy ratio A is computed and exhibits a discontinuity at the transition.}
\label{DV.fig}
\end{center}
\end{figure}

\begin{figure}[!htb]
\begin{center}
\includegraphics[width=75mm]{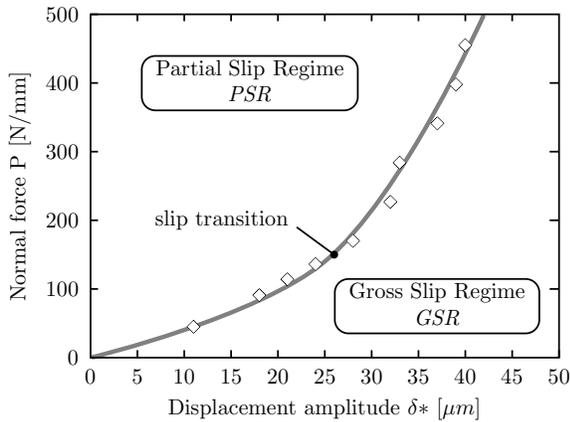}
\caption{Experimental determination of the sliding transition, cylinder/plane contact, 2024T351 vs 7075T6, R=49mm.}
\label{sliding_transition.fig}
\end{center}
\end{figure}

To determine the fretting regimes, variable displacement method \cite{Voisin1992}, has been applied for different values of the normal load. In each test, the normal load is kept constant. The relative displacement start to a very low value, imposing partial slip (with $Q^*<\mu_t P$). When stabilised conditions are reached, $\delta^*$ is increased and maintained constant until new stabilisation is reach. $\delta^*$ is increased this way, step by step, until the contact has overcome the sliding transition conditions (i.e. $Q^*=\mu_t P$). Precise value of the transition is then calculated by computing the value of the energy sliding criterion $A=E_d/E_t$ ; with $E_d$ the dissipated energy of the corresponding cycle and $E_t$ the total energy \cite{Fouvry1996,Fouvry2002}. In the cylinder/plane configuration, it has been shown that the sliding transition is associated to a discontinuity of ratio A (see figure \ref{DV.fig}).
Figure \ref{sliding_transition.fig} present the sliding transition obtained with the variable displacement method, for a cylinder/plane contact of Al2024T351 versus Al7075T6. On this figure, results are presented versus the relative displacement amplitude $\delta^*$. However, it does not appear as the more convenient representation when different series of results are compared. Indeed, the value of $\delta^*$ depends on the fretting setup compliance \cite{Fouvry2002}. Since we will use 3 different fretting pads and different contact compliances, the displacement amplitude is adequately substituted by the linear tangential force amplitude Q$^*$(N/mm). Note that the tangential force is not affected by the apparatus compliance. In the next, except if specified, results will be presented using the tangential force.
\subsection{Coefficient of friction in partial slip}\label{local_friction.ssc}
\begin{figure}[!hbt]
\begin{center}
\includegraphics[width=75mm]{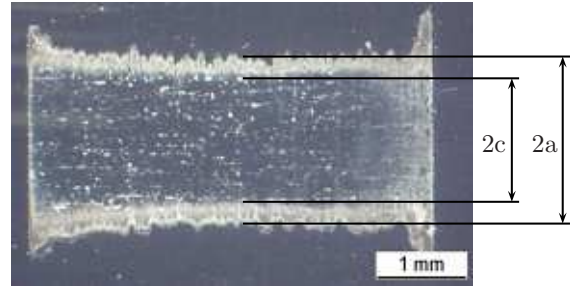}
\caption{Experimental scar after a fretting test, Fn=1400N and Ft=500N. Definition of stick and slip zones width.}
\label{scar_R1.fig}
\end{center}
\end{figure}

\begin{figure}[!hbt]
\begin{center}
\includegraphics[width=75mm]{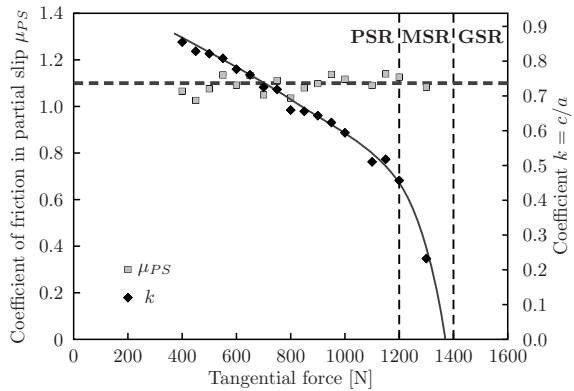}
\caption{Determination of the friction law in partial slip regime, $F_N=1400N$.}
\label{local_friction.fig}
\end{center}
\end{figure}

Under partial slip conditions, slip takes place in two symmetrical regions $a>|x|\geq c$ which surround a central stick zone $|x|<c$. In this central zone, from Coulomb's law we have $q(x)\leq \mu_t \times p(x)$. Because of the stick zone, the friction coefficient obtained at the sliding transition cannot not used. Instead, the friction coefficient in partial slip regime $\mu_{PS}$ can be determined by the investigation of the scars, after the fretting tests. From Mindlin cited by Jonhson \cite{Johnson1985} we have: 
\begin{equation}
c=a\left(1-\frac{Q^*}{\mu_{PS}\,P}\right)^{1/2}
\label{c_expr.eqn}
\end{equation}
From which:
\begin{equation}
\mu_{PS}=\frac{1}{1-\left(\dfrac{c}{a}\right)^2}{\frac{Q^*}{P}}
\label{muL_expr.eqn}
\end{equation}
From examination of fretting scars, the contact width $2a$ and the stick zone width $2c$ (defined on figure~\ref{scar_R1.fig}) are measured for a large range of displacement values and the friction coefficient in partial slip regime is calculated in each case. This was done for a constant normal force $F_N=1400N$ ($P=318N/mm$) and result is assumed to be true wathever the value of P is for the investigated experimental conditions. Figure~\ref{local_friction.fig} summarise the results obtained by this methodology.
The evolution toward gross slip as displacement (and so tangential force) increase is well observed. Calculations of the friction coefficient in partial slip regime with equation (\ref{muL_expr.eqn}) show clearly a constant evolution, very close to the value at the transition. Then we can postulate that the coefficient of friction under studied partial slip conditions, is identical to the value defined at the transition.
\begin{equation}
\mu_{PS}(k,Q^*,P)=\mu_t\cong 1.1
\label{muL_eq_mut.eqn}
\end{equation}
\section{Cracking analysis in partial slip conditions}\label{crack.sec}
The experimental objective in this part is to determine as finely as possible for each studied normal force, the critical tangential force associated to the crack initiation after $5.10^4$ cycles under stabilised partial slip conditions.
\subsection{Methodology}
\begin{figure}[!hbt]
\begin{center}
\includegraphics[width=75mm]{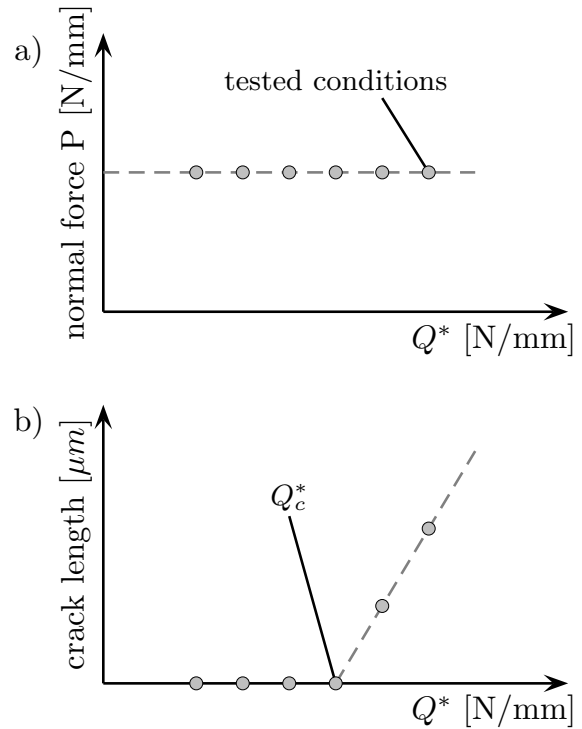}
\caption{Schematics of the crack nucleation threshold determination: a) fretting test conditions ; b) results in terms of crack length.}
\label{methodology.fig}
\end{center}
\end{figure}

For each normal load, kept constant during a test, several fretting tests were carried out as detailed in~\ref{fretting_exp.ssc}, maintaining a constant tangential force (Q$^*$) by monitoring the displacement amplitude ($\delta^*$). The critical value (Q$^*_c$) is precisely determined thanks to cross section observations of specimens cut after fretting solicitation. Cracking is assessed by optical inspection, the smallest observable crack being around $5\mu m$ long. To bracket the crack nucleation threshold, dichotomy is applied to select the values of Q$^*$ \cite{Fouvry1996}. About ten test are required to reach a 5N accuracy in the determination of Q$^*_c$. Measurement of the crack lengths allows to extrapolate Q$^*_c$ even if dispersion can sometimes disturb the method. This methodology is illustrated in figure~\ref{methodology.fig}.
\subsection{Results with a smooth contact}\label{smooth_res.ssc}
\begin{figure}[!hbt]
\begin{center}
\includegraphics[width=75mm]{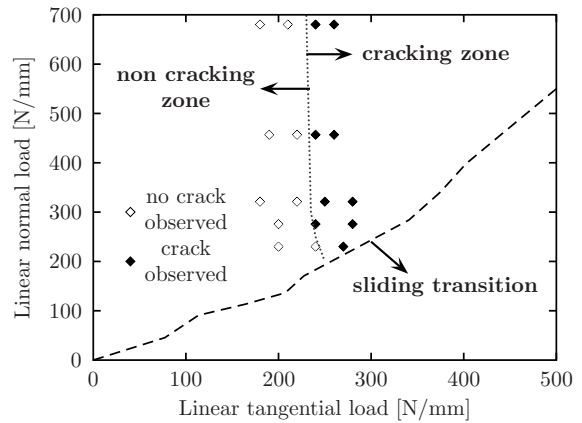}
\caption{Experimental fretting crack nucleation boundary, cylinder/plane smooth contact (R=49mm).}
\label{crack_boundary.fig}
\end{center}
\end{figure}

The described methodology is applied with R1 counter body for 5 normal forces and crack nucleation boundary is determined (figure \ref{crack_boundary.fig}). In addition, Several other conclusions can be derived from experimental observations. First of all, the crack nucleation is always located near the trailing edge of the contact, which is in agreement with former fretting studies in partial slip regimes. When crack initiation is detected, the crack angle is measured and is found to be lower than 35 degrees for all the loading conditions investigated. Referring to the work of Lamacq \cite{Lamacq1997} and Dubourg and Lamacq\cite{Dubourg2000}, cracks seems to growth under mode II by maximal shear amplitude.
Figure \ref{crack_boundary.fig} shows that the threshold tangential load is found independent of the normal force, giving: 
\begin{equation}
Q_c^*(P)=Q_c^*\cong 240N/mm
\end{equation}
\section{Roughness Impact}\label{rough.sec}
In this section, results with rough surfaces are presented and compared. As expected, significant gaps are observed between each crack nucleation boundary. The effective contact area concept is introduced in order to account for the roughness. Finally, a new representation is introduced which rationalises the different crack boundaries obtained for different roughnesses.
\subsection{Results with rough contacts}
\begin{figure}[!hbt]
\begin{center}
\includegraphics[width=75mm]{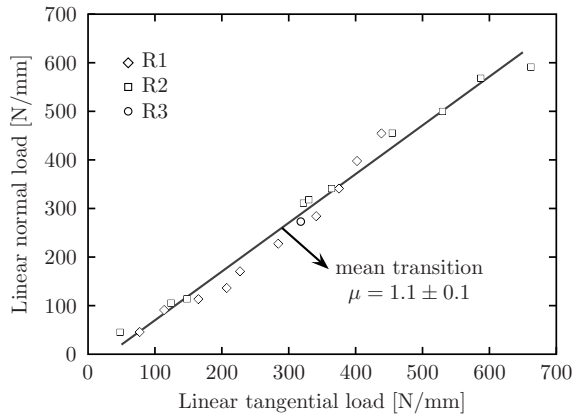}
\caption{Comparison of sliding transitions obtained with smooth and rough cylinder.}
\label{rough_transition.fig}
\end{center}
\end{figure}

\begin{figure}[!hbt]
\begin{center}
\includegraphics[width=75mm]{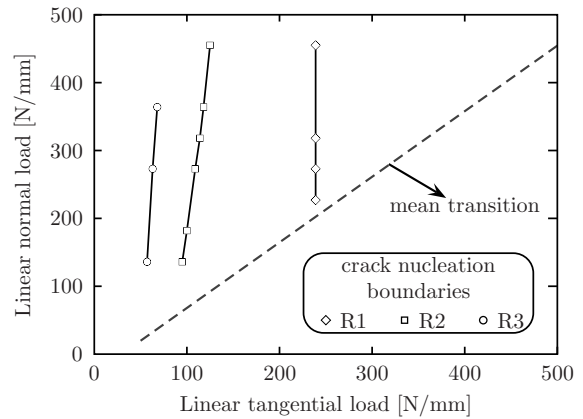}
\caption{Experimental determination of crack nucleation boundaries with three different roughness.}
\label{crack_boundaries.fig}
\end{center}
\end{figure}

\begin{figure}[!hbt]
\begin{center}
\includegraphics[width=75mm]{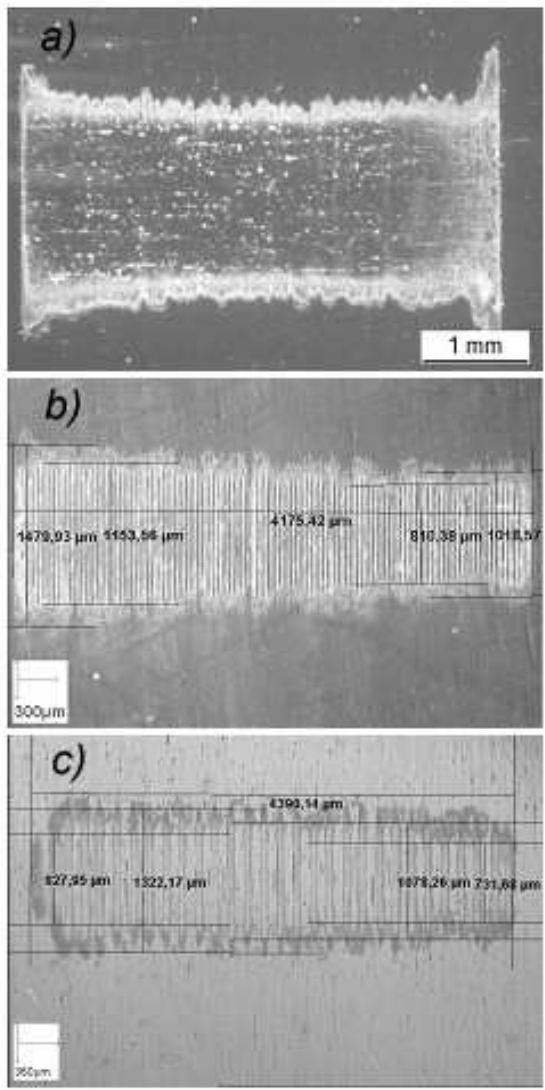}
\caption{Examples of optical micrography of specimen surfaces after a fretting test. a) pad R1, $F_N=1400N$ and $F_T=500N$ b) pad R2, $F_N=800N$ and $F_T=400N$ c) pad R3, $F_N=600N$ and $F_T=450N$.}
\label{fretting_scars.fig}
\end{center}
\end{figure}

Mean and total values of fretting pads roughness were presented in table \ref{roughness.tab}. The same methods as used in paragraph \ref{exp_slid.sss}, for determining the sliding transition and in paragraph \ref{smooth_res.ssc} for establishing the crack nucleation boundary, were used with the fretting pads of roughness R2 and R3. The corresponding results are shown in figures \ref{rough_transition.fig} and \ref{crack_boundaries.fig}. On fig.~\ref{rough_transition.fig} it can be seen that the sliding transition does not depends on the roughness value. This could be expected since gross slip condition is imposed at the transition, which induce wear in the contact. Roughness is then blurred in a few cycles; and the contact interface tends to a unique structure.
On the contrary, the crack initiation, under stabilised partial slip condition, strongly depends on the surface roughness (figure~\ref{crack_boundaries.fig}). As expected, a higher roughness leads to a lower tangential threshold load. Fig~\ref{crack_boundaries.fig} also shows a slope in the experimental boundaries obtained with R2 and R3. This influence of the normal force on the tangential threshold load was not observed with R1.
Those differences in tangential load threshold can be explained in considering intrinsic contact parameters. Indeed, the observation of fretting scars (figure~\ref{fretting_scars.fig}), clearly shows that for identical loading conditions, the fretting pad surface in contact varies with the surface roughness. This implies different loading conditions inside the contact, depending on the area effectively involved in the contact. 
\subsection{Definition of the effective contact area}
\begin{figure}[!hbt]
\begin{center}
\includegraphics[width=75mm]{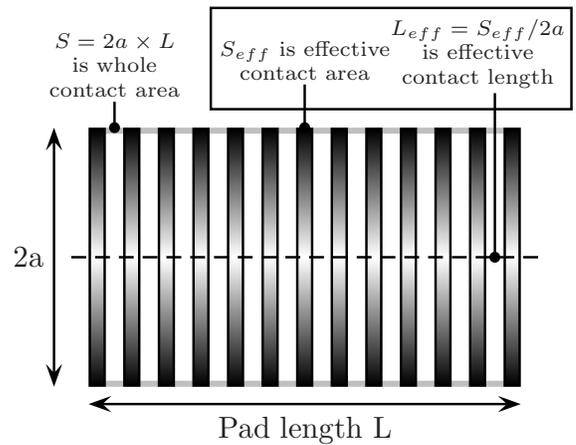}
\caption{Definition of the effective contact area $S_{eff}$ (see the text for details).}
\label{Seff.fig}
\end{center}
\end{figure}

\begin{figure}[!hbt]
\begin{center}
\includegraphics[width=75mm]{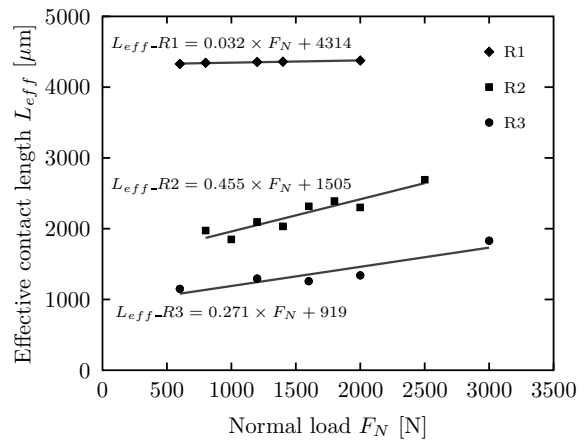}
\caption{Experimental determination of $L_{eff}$ evolution with normal force (pad length~:~$L=4.4mm$).}
\label{Leff.fig}
\end{center}
\end{figure}

First developed by Greenwood \cite{Greenwood1967}, the effective contact area $S_{eff}$ can be simplified in our case. Considering the polished surfaces of the specimens and the unidimensional fretting pad roughness, $S_{eff}$ is easily defined as the sum of the micro-rectangles of the pad indentation. This definition can be one more time simplified by $2a \times L_{eff}$ where $L_{eff}$ is the sum of the micro-length of indentation on the medium line of the contact (figure~\ref{Seff.fig}). This rather simplistic definition is only valid for well aligned contact.
Next step is to quantify the influence of $F_N$ and $F_T$ on $L_{eff}$. Effective contact lengths were measured from optical images of fretting scars after the tests. Results clearly show a negligible influence of tangential load so that $L_{eff}$ appears mainly monitored by the normal load. A linear dependence of $S_{eff}$ with $F_N$ is observed~; the quasi horizontal curve obtained for R1 shows that $S_{eff}=S$ in the investigated loading range (figure~\ref{Leff.fig}).
\subsection{Unification of the fretting crack nucleation boundary}
\begin{figure}[!hbt]
\begin{center}
\includegraphics[width=75mm]{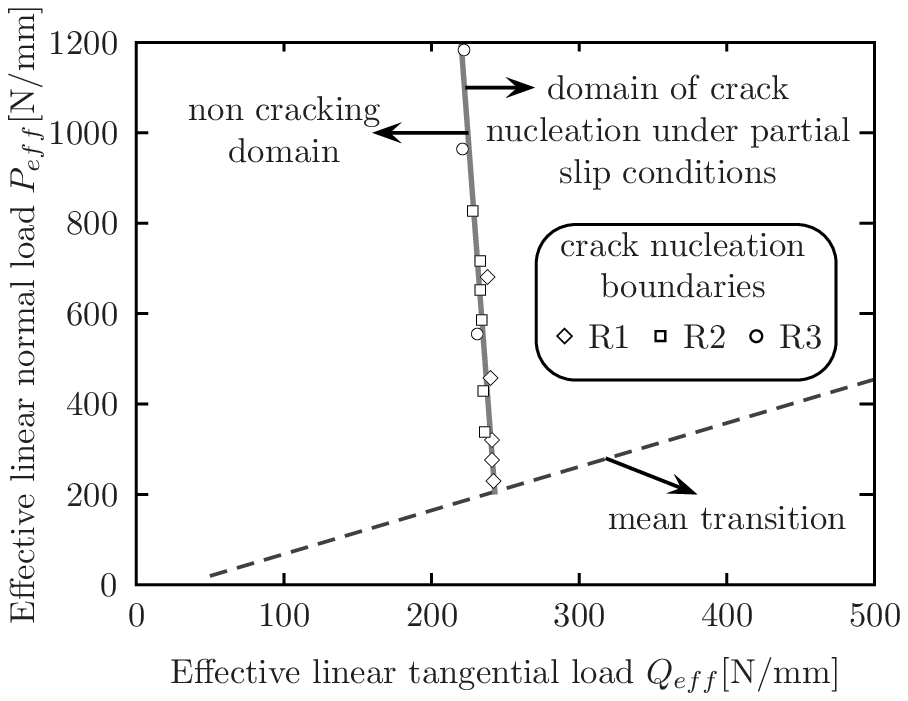}
\caption{Experimental fretting crack nucleation boundaries, with various pad roughness, plotted in $P_{eff}/Q_{eff}$ representation.}
\label{unification.fig}
\end{center}
\end{figure}

Two new loading parameters $P_{eff}$ and $Q_{eff}$, respectively the effective normal load and the effective tangential load are introduced~: 
\begin{equation}
P_{eff}=\frac{F_N}{L_{eff}}\mbox{  and  }Q_{eff}=\frac{F_T}{L_{eff}}
\label{PQeff.eqn}
\end{equation}
Fig. \ref{Leff.fig} gives the evolution of $L_{eff}$ with $F_N$. Assuming a negligible impact of the tangential force on $L_{eff}$, the values of $P_{eff}$ and $Q_{eff}$ can be computed. All the fretting tests defining the three nucleation boundaries can then be plotted in terms of the intrinsic contact loading parameters $P_{eff}$ and $Q_{eff}$. The corresponding representation is presented on figure~\ref{unification.fig}.
On this figure, compared to the fig. \ref{crack_boundaries.fig}, the crack nucleation boundary under partial slip conditions appears independant of the roughness value (for the experimental conditions investigated). Accessing intrinsic contact loading parameter was successful in correlating the three experimental boundaries. One point further reinforce this correlation: in fig. \ref{crack_boundaries.fig}, the slope with R2 is larger than with R3 even if the roughness is lower. Figure \ref{Leff.fig} shows that the slope in the $L_{eff}$ evolution with $F_N$ is larger for R2 than for R3. This result might seem incoherent with the fact that zero slope is expected for a perfectly smooth contact. However it could be explained by a different response of roughness to indentation induced by the quite different shapes of the roughness profiles shown on fig. \ref{profiles.fig}. This outlines the difficulty to rationalise a surface morphology by only considering the two parameters R$_a$ and R$_t$. In any case, the larger slope observed for the $L_{eff}\_R2(F_N)$ curve perfectly accounts for the good correlation observed on figure \ref{unification.fig} and highlights the physical reliability of the approach.\\
To summarise fig. \ref{unification.fig}, it is shown here that the crack initiation of the 2024 alloy studied depends on the real local stress state ($P_{eff},Q_{eff}$), which cannot be calculated without considering the real contact length $L_{eff}$, in case of a rough contact. It is also shown here that considering the effective loading parameters have removed the effect of P observed with the two rough contacts on fig. \ref{crack_boundaries.fig}. This effect may thus be related to the material response versus the indentation of the asperities of the counter body.
\section{Fretting crack initiation prediction with SWT parameter}\label{SWT.sec}
To predict the crack nucleation risk and formalise the previous experiments, the SWT multiaxial fatigue criterion is implemented and combined with a Mindlin description of the cylinder/plane contact configuration. Explicit formulation will provide a physical justification to a non dependence of the crack initiation boundary with respect to the normal load. Size effect is also introduced and implemented. It should be noted that alternative approaches like combining classical fracture mechanics and notch fatigue can also be used to tackle this problem. Such an approach was recently carried out by Ciavarella \cite{ciavarella2003}.
\subsection{Theoretical background}
The SWT critical plane approach was first used to tackle fretting problems by Szolwinski and Farris \cite{Szolwinski1996}. They used a fatigue approach developed by Smith et al. \cite{Smith1970} which was developed to account for a mean stress or strain effect in classical fatigue. According to the Szolwinski and Farris approach, the crack initiation in fretting occurs on the plane where the product of the normal strain amplitude $\epsilon _a$, and the maximum normal stress $\sigma _{max}$ is maximum. The SWT parameter $\Gamma$ can then be expressed as: 
\begin{equation}
\Gamma =\sigma _{max}\times \epsilon _a=\frac{(\sigma _f')^2}{E}(2N)^{2b'}+\sigma _f'\epsilon _f'(2N)^{b'+c'}
\label{SWT.eqn}
\end{equation}
where $\sigma _f'$ is the fatigue strength coefficient, $b'$ is the fatigue strain exponent, $\epsilon _f'$ is the fatigue ductility coefficient, c' is the fatigue ductility exponent and $N$ the number of reversal to failure. These constants are also given by Szolwinski et al. \cite{Szolwinski1998} and listed in table~\ref{swt_par.tab}.\\
\begin{table}[!hbt]
\caption{SWT parameters for Al 2024T351 from \cite{Szolwinski1998}}
\label{swt_par.tab}
\begin{tabular}{cccc}
	$\sigma _f'$(MPa) & $b'$ & $\epsilon _f'$ & $c'$ \\
	\hline
	714 & -0.078 & 0.166 & -0.538 \\
\end{tabular} 
\end{table}
\noindent To normalise the crack nucleation risk, the following scalar parameter $d_{SWT}$ can be introduced~:
\begin{equation}
d_{SWT}=\frac{max(\sigma _{max}\times \epsilon _a)}{\frac{(\sigma _f')^2}{E}(2N)^{2b'}+\sigma _f'\epsilon _f'(2N)^{b'+c'}}
\label{dswt.eqn}
\end{equation}
If $d_{SWT}$ is greater than or equal to 1, cracking is likely to occur. If $d_{SWT}$ is less than 1, there is no risk of cracking. By expressing analytically $\sigma _{max}$ and $\epsilon _a$, a literal expression of the cracking risk $d _{SWT}$ can be obtained. Fouvry \cite{Fouvry2002} and Fridrici \cite{Fridrici2002} showed that the SWT multiaxial fatigue criteria application on fretting contact gives a maximum crack initiation risk a the trailing edge of the contact. This is in agreement with our experimental observations. We can therefore express the surface stresses at the contact border in partial slip with Mindlin theory. Assuming an elastic plane strain 2D state, we use the general equation \cite{Hills1994}~: 
\begin{eqnarray}
\sigma _{ij}(x,y)&=&p_0\left(\frac{\sigma _{ij}^n\left(\frac xa ,\frac ya\right)}{p_0}
	\right)+\mu p_0\left(\frac{\sigma _{ij}^t\left(\frac xa ,\frac ya\right)}{\mu p_0}\right)\nonumber\\
	&\ &-\mu p_0\frac ca\left(\frac{\sigma _{ij}^t\left(\frac xc ,\frac yc\right)}{\mu p_0}\right)
\label{PSR_stress.eqn}
\end{eqnarray}
Where $\sigma _{ij}^n$ and $\sigma _{ij}^t$ denotes stresses in a cylinder plane contact submitted to normal load and tangential load respectively. Note that the subscript $ij$ can denote any spatial direction. At the trailing edge of the contact (at $x=a$ and $y=0$) we have~: 
\begin{eqnarray}
\left\{
	\begin{array}{l}
		\sigma _{xx}^n(1,0)=0 \\
		\sigma _{yy}^n(1,0)=0 \\
		\sigma _{zz}^n(1,0)=0 \\
		\sigma _{xy}^n(1,0)=0
	\end{array}
\right.
\textrm{, }
\left\{
	\begin{array}{l}
		\sigma _{xx}^t(1,0)=2\mu p_0     \\
		\sigma _{yy}^t(1,0)=0            \\   
		\sigma _{zz}^t(1,0)=2\nu \mu p_0 \\
		\sigma _{xy}^t(1,0)=0		   
	\end{array}
\right.&\ \nonumber\\
\textrm{and}
\left\{
	\begin{array}{l}
		\sigma _{xx}^t(\frac ac,0)=2\mu p_0 \frac ac -2\mu p_0 \sqrt{(\frac ac)^2-1} \\
		\sigma _{yy}^t(\frac ac,0)=0 	\\
		\sigma _{zz}^t(\frac ac,0)=\nu \times \sigma _{xx}^t(\frac ac,0) \\
		\sigma _{xy}^t(\frac ac,0)=0
	\end{array}
\right.&
\end{eqnarray}
which gives~:
\begin{equation}
\sigma _{xx}(a,0)=2\mu p_0\sqrt{1-\left(\frac ca\right)^2}
\label{sigxxa0.eqn}
\end{equation}
This equation may be written in a more convenient form by combining it with (\ref{c_expr.eqn}) to give~:
\begin{equation}
\sigma _{xx}(a,0)=2\mu p_0\sqrt{\frac{Q^*}{\mu _{PS}P}}
\label{sigxxa0bis.eqn}
\end{equation}
In cylinder/plane configuration we have $p_0=\left(\frac{PE^*}{\pi R}\right)^{1/2}$ and experimental analysis (see \ref{local_friction.ssc}) showed that $\mu _{PS}=\mu =1.1$. Combining these elements we get~: 
\begin{equation}
\sigma _{xx}(a,0)=2\sqrt{\frac{\mu E^*Q^*}{\pi R}}
\label{sigxxa0ter.eqn}
\end{equation}
The stress state at the trailing edge of the contact can be considered as quasi uniaxial. 
From Hook's law we deduce at point (a,0)~: 
\begin{equation}
\epsilon _{xx}=\frac 1E[(1-\nu ^2)\sigma _{xx}-\nu (1+\nu )\sigma _{zz}]= 
	\frac 1E[(1-\nu ^2)-\nu ^2(1+\nu )]\sigma _{xx}
\label{epsxx.eqn}
\end{equation}
On the loading path of a fretting cycle, we have at point (a,0)~: 
\begin{equation}
\epsilon _a=\frac {\Delta E}{2}=\frac {\epsilon _{xx}-(-\epsilon _{xx})}{2}=\epsilon _{xx}
\label{epsa.eqn}
\end{equation}
The SWT parameter is finally simplified to the following expression~: 
\begin{equation}
\Gamma =\frac {1-2\nu ^2 -\nu ^3}{E}\sigma _{xx}^2
\label{swt_expr.eqn}
\end{equation}
We consider the critical crack initiation limit ($d_SWT=1$) so that the equivalent critical crack tensile stress, with respect to SWT criterion, is expressed as~: 
\begin{equation}
{\sigma _{xx}^2}_c^{Q*}=\frac {(\sigma _f')^2(2N)^{2b'}+\sigma _f'\epsilon _f'(2N)^{b'+c'}}{1-2\nu ^2 -\nu ^3}
\label{epsxx2.eqn}
\end{equation}
Finally from equation (\ref{sigxxa0ter.eqn}) we determine a critical linear tangential force amplitude associated to the crack initiation condition~: 
\vspace{0.5cm}
\begin{equation}
Q_c^*=\frac {\pi R}{4\mu E*}\frac {(\sigma _f')^2(2N)^{2b'}+\sigma _f'\epsilon _f'(2N)^{b'+c'}}{1-2\nu ^2 -\nu ^3}
\label{Qc*.eqn}
\end{equation}
The critical linear tangential force amplitude is proportional to the cylinder radius, a function of the elastic and fatigue properties of the material and inversely proportional to the friction coefficient. It must be pointed out that $Q_c^*$ does not depend on the normal force which account for the observed negligible impact of the pressure field on crack nucleation. Performing a numerical application leads to $Q_c^*=106 N/mm$, value to be compared with the experimental threshold of 240 N/mm. This difference could be related to the large stress/strain gradients which exists in and below the contact region.\\ 
Although this approach gives a theoretical justification of the normal load non dependance, it cannot predict the exact threshold tangential load. Indeed, fretting crack initiation behaviour may also depend on the severity of the stress/strain gradients, and not only on the maximum values. First introduced by Fouvry et al. \cite{Fouvry1998,Fouvry2002}, the crack nucleation process volume was successively adapted by Araujo and Nowell \cite{Araujo2002} and more recently used by Naboulsi and Mall \cite{Naboulsi2003}. In this approach, the computation of a representative SWT critical plane parameter must be conducted from an averaged stress strain state, defined over an intrinsic processing material volume. Hence, the stress gradient effect is indirectly controlled through the averaging process.
\subsection{Crack nucleation process volume approach}
\begin{figure}[!hbt]
\begin{center}
\includegraphics[width=75mm]{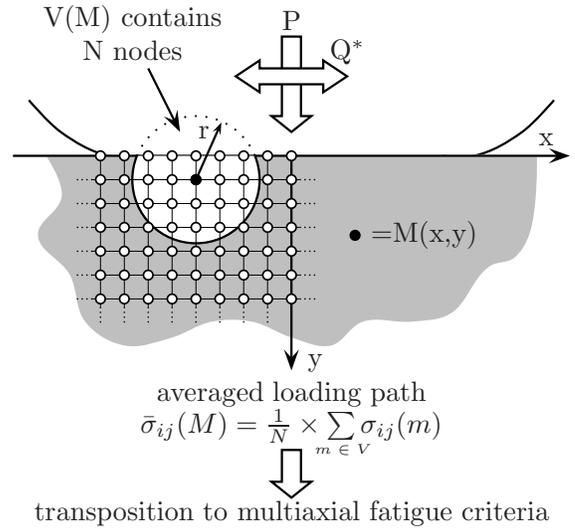}
\caption{Schematic representation of the radial shape process volume averaging.}
\label{process_volume.fig}
\end{center}
\end{figure}

\begin{figure}[!hbt]
\begin{center}
\includegraphics[width=75mm]{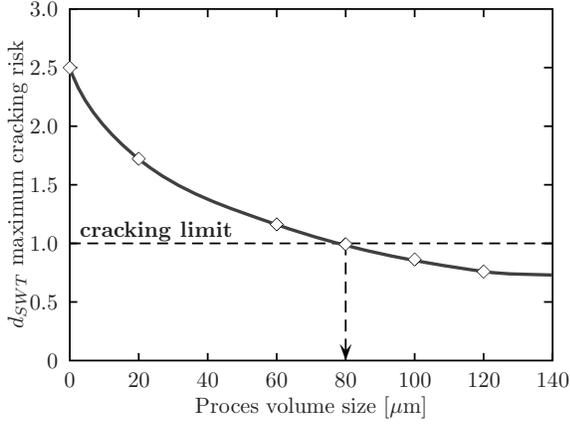}
\caption{Application of the averaging method in calculation of the cracking risk at surface with the SWT parameter.}
\label{size_effect.fig}
\end{center}
\end{figure}

\begin{figure}[!hbt]
\begin{center}
\includegraphics[width=75mm]{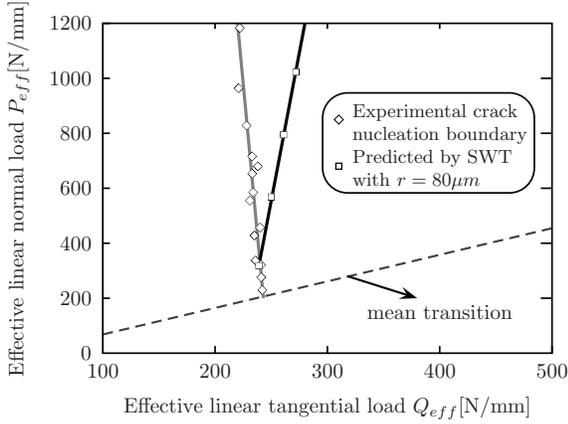}
\caption{Comparison between experimental and predicted crack nucleation boundary with a $80\mu m$ size effect. Fretting contact: 
Al2024T351 plane versus Al7075T6 cylinder pad (R=49mm).}
\label{boundary_ana_exp.fig}
\end{center}
\end{figure}

To compute the SWT parameter using the process volume approach, a cross section is discretised and partial slip stresses and strains corresponding to the experimental conditions, are computed from Mindlin solutions. Process volume averaging is then applied considering a radial shape volume as shown on figure \ref{process_volume.fig}.
The SWT crack initiation parameter is then computed from these averaged values which gives a mean SWT value on the process volume. As discussed in \cite{Naboulsi2003}, SWT could be computed on each point, the averaging procedure occurring thereafter ; but the result would not be very different and the first method, which is faster, has more physical meaning. From these computations, the critical locations where cracking risk is maximum, is located at the contact surface in the vicinity of the trailing edge. This matches the experimental observations seen in section \ref{crack.sec}. However if the crack location at the trailing edge is well predicted, the SWT parameter does not predict the observed crack initiation angle ($\leq$35 degrees). All the values are closed to 90 degrees, perpendicular to the maximum principal stress. Indeed, mainly dependent on the tensile stress and strain state, the SWT parameter is unable to describe the shear mechanisms which presently control the crack nucleation process. The process volume has been identified for a linear effective force of 380 N/mm. Figure~\ref{size_effect.fig} displays the surface distribution of the cracking risk $d_{SWT}$ defined in equation (\ref{dswt.eqn}), for different averaging length scale values. Experimental correlation with the crack nucleation boundary is obtained for a $r=80\mu m$ process volume radius. This value match very well the mean grain size radius of $75 \mu m$, and consequently confirms a correlation between the mechanical crack nucleation process volume and the microstructure.
With the process volume radius fixed, by varying normal load in the computation analysis, one can draw the analytic crack initiation boundary predicted with SWT parameter. This is compared to the experimental boundary on fig.~\ref{boundary_ana_exp.fig} and plot in the effective loads representation.
\subsection{discussion}
Obviously, figure~\ref{boundary_ana_exp.fig} confirms that SWT computations with a $r=80\mu m$ process volume radius allows to predict the studied 240 N/mm linear normal loading. Although a weak influence of the normal load is observed, it is nevertheless deduced that the application of a singular process volume radius is ineffective to quantify the pressure impact. Indeed, in the SWT formulation, pressure effect have tendency to reduce the cracking risk which is based on tensile stress, although the contrary is observed experimentally. Thus the physical meaning of the SWT parameter to predict the 2024T351 fretting crack initiation appears discutable. Other criteria, more representative of the physical behaviour of the studied alloy, should be studied involving elasto-plastic analyses of the stress/strain loading path.
\subsection{Identification of a safe crack nucleation process volume}
The results presented above show that analytical calculations coupled with the process volume approach and the SWT criterion are limited for predicting the experimental crack nucleation boundary. However, due to the simplicity of its formulation, it is widely employed in industrial component conception. An alternative method to reach a safe prediction for crack initiation in contact submitted to fretting loading, is presently introduced.\\
First, the contact pressure range must be defined. Then the process volume is evaluated for both $P_{min}$ and $P_{max}$. Assuming a monotonous evolution between $P_{min}$ and $P_{max}$, the safe crack nucleation process volume $r_s$ is defined as~:
\begin{equation}
r_s=Min[r(P_{max}),r(P_{min})]
\label{rsafe.eqn}
\end{equation}
\begin{figure}[!hbt]
\begin{center}
\includegraphics[width=75mm]{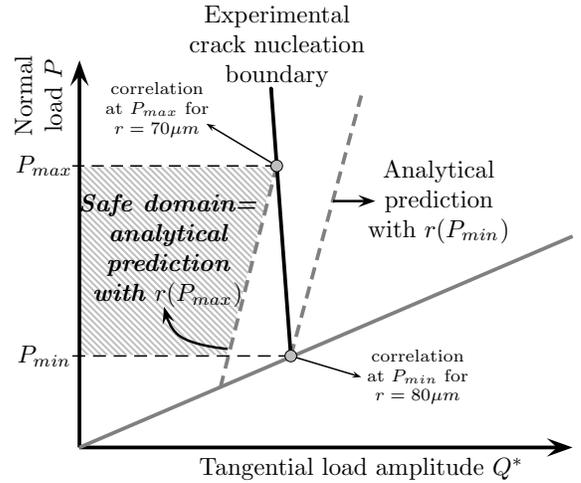}
\caption{Definition of the safety domain predicted by analytical computations of SWT parameter with the safe crack nucleation process 
volume $r_s$.}
\label{safe_prediction.fig}
\end{center}
\end{figure}
In the present case, a positive slope with the pressure increase has been identified. Therefore we can directly assume that the safe crack nucleation process volume correspond to the process volume defined for $P_{max}$, the maximum pressure loading. Setting $P_{max}$ to 1000N/mm leads to a critical value of $r=70\mu m$ which can be used to define the safety domain (see figure~\ref{safe_prediction.fig}). It is important to mention that the obtained process volume is very close to the value previously defined for the minimal pressure. Again, it confirms the stability of the approach and support the idea of a reliable grain size averaging approach.
\section{Conclusion}\label{concl.sec}
Fretting crack initiation under partial slip conditions is investigated through an experimental approach and the effect of roughness is studied quantitatively. Aluminium alloys contact, under cylinder/plane configuration is studied in order to tackle crack initiation in riveted assemblies. Three fretting pads are used with tree different surface qualities, with a unidimensional roughness morphology.

First, the running condition fretting map is determined and the friction law is quantified. the friction coefficient in partial slip is demonstrated to be equal to the friction coefficient at the transition $\mu_t=1.1$. The sliding transition appears to be independent of the roughness value, which could be explained by a rapid blurring of the incipient surface roughness under gross slip conditions.

Further work has been undertaken to determine the experimental crack nucleation boundaries with respect to the surface roughness. Results have shown that a critical tangential load could be found with a smooth contact, although a weak pressure effect is observed for the two rough contacts. As expected, crack initiation appears very sensitive to the surface quality, a higher roughness leading to a lower  value of the tangential force needed for crack initiation. A contact effective area formulation is introduced in order to access the intrinsic contact loading parameters. Linear normal and tangential effective loads $P_{eff}$ and $Q_{eff}$ are computed and a new representation is drawn to describe the crack nucleation boundaries. This is successfully applied to correlate experimental results and various crack nucleation boundaries are unified independently of the surface quality leading to $Q^*_{effc}=240 N/mm$.

An analytical justification of the weak dependence of $Q_c^*$ to the normal load for a smooth contact has been conducted. SWT criterion has been formulated in order to predict the crack nucleation threshold. The non influence of the normal load on crack nucleation threshold is justified although the numerical application appears to be on the safe side. A radial process volume approach is then introduced to take into account the severe stress/strain gradient in the contact. A global correlation is found with the experimental results with a process volume diameter which match the mean grain radius of 75~$\mu m$ and thus, this effect could be correlated to a stress/strain microstructure accommodation. In spite of this, the pressure effect on the crack nucleation boundary, is not in agreement with experimental results. To avoid a large underestimation of the cracking risk, a safe crack nucleation process volume identification approach is introduced. It consists in defining the smallest volume through the investigated pressure and contact size ranges.

However, some critical aspects like the pressure loading impact or the incipient crack orientation suggests that other fatigue approaches must be considered and elasto-plastic stress/strain analysis are required.

\section*{Acknowledgements}
This research is supported by the Rhone Alpes Material Federation, Pechiney and EADS Industries.

\bibliography{bibliography}

\end{document}